\documentclass[12pt,a4paper]{article}
\usepackage[dvips]{graphicx}

\newcommand{\beq}{\begin{equation}}
\newcommand{\eeq}{\end{equation}}
\newcommand{\bea}{\begin{eqnarray}}
\newcommand{\eea}{\end{eqnarray}}
\begin{document}
\thispagestyle{empty}
\vspace*{-15mm}
%----------
\baselineskip 10pt
\begin{flushright}
\begin{tabular}{l}
{\bf OCHA-PP-229}\\
{\bf July 2004}\\
{\bf hep-th/0407118}
\end{tabular}
\end{flushright}
\baselineskip 24pt 
\vglue 10mm 
%%%%%%%%%%%%%%%%%%%%%%%%%%%%%%%%%%%%%%%%%%%%%% 
%                Title 
%%%%%%%%%%%%%%%%%%%%%%%%%%%%%%%%%%%%%%%%%%%%%%
\begin{center}
{\LARGE\bf{Singular Gauge Transformation in Non-Commutative $U(2)$ Gauge Theory
}}
\vspace{7mm}\\
\baselineskip 18pt 
{\bf
Yuko KOBASHI, Akio SUGAMOTO}
\vspace{2mm}\\
{\it 
 Department of Physics, Ochanomizu University, \\
 2-1-1, Otsuka, Bunkyo-ku, Tokyo 112-8610, Japan
}\\
\vspace{10mm}
\end{center}
%%%%%%%%%%%%%%%%%%%%%%%%%%%%%%%%%%%%%%%%
%%%%%                              %%%%%
%%%%%          Abstract            %%%%%
%%%%%                              %%%%%
%%%%%%%%%%%%%%%%%%%%%%%%%%%%%%%%%%%%%%%%
\begin{center}
{\bf Abstract}\\[7mm]
\begin{minipage}{14cm}
\baselineskip 16pt
\noindent
%%%%%----------------------------------
A method developed by Polychronakos to study singular gauge
transformations in 1+2 dimensional non-commutative Chern-Simons gauge
theory is generalized from $U(1)$ group to $U(2)$ group. The method
clarifies the singular behavior of topologically non-trivial gauge transformations 
in non-commutative gauge theory, which appears when the gauge transformations 
are viewed from the commutative gauge theory equivalent to the commutative theory.
%%%%%----------------------------------  
\end{minipage}
\end{center}
%%%----------------------------------  
%%%
%%%
%%%   Main body of the paper
%%%
%%%
%%%----------------------------------  
\newpage
\baselineskip 18pt 
\def\thefootnote{\fnsymbol{footnote}}
\setcounter{footnote}{0}
%%%%%%%%%--- 	Intro --- %%%%%%%%%%%%%%%%%%%
\section{Introduction}
%%%%%%%%%%%%%%%%%%%%%%%%%%%
%%%------------
%%%
Quantum Hall effect (QHE) is the phenomenon which occurs in the 1+2 dimensional  
electron fluid under a strong magnetic field applied from outside. 
It is interesting to note that as is pointed out by Susskind, QHE can be equivalently 
described using a non-commutative Chern-Simons (CS) theory or a Matrix model of 
CS-type~\cite{QHS}.
%---

%---
Therefore, the treatment of QHE becomes closer to that of non-commutative gauge 
theory and string theory in particle physics. In these equivalent descriptions, 
Laughlin theory with a filling fraction $\frac{1}{n}$ (n=integer) becomes 
non-commutative CS theory with a CS factor (a coefficient of CS action) being n.
%---

%---
In the original system of electron fluid, the fluid becomes incompressible due to 
the Pauli principle, so that the area occupied by electrons is conserved dynamically. 
The freedom to change the coordinate system of fluid, by preserving the area, 
is called Area Preserving Diffeomorphism(APD). The symmetry relating to APD is 
transferred to the non-commutative $U(1)$ gauge transformation in the non-commutative 
CS theory, while it becomes in the matrix model of CS-type to be the unitary 
$U(\infty)$ transformation of the matrix valued coordinates describing electrons.
%---

%---
Quasi-particles and quasi-holes appear as the surplus and deficit of
area in the fluid system, or the singularity of the APD.  Equivalently,
 they are described by singular gauge transformations in the non-commutative CS
theory, since APD and gauge transformation are equivalent symmetries in different 
descriptions.  In this manner, we can understand the importance of  studying singular 
gauge transformations in non-commutative gauge theory.
%---

%---
Recently, in order to study the exciton state having both quasi-electron
and quasi-hole, the usual one matrix model is found not useful in order
to give the exciton state, but two matrix model is to be introduced~\cite{QHEX}.
%---

%---
  In this two matrix model, exciton solution is obtained and its dispersion
relation is estimated.  The estimated dispersion relation shows a stable
point at which the distance between hole and electron takes a fixed
value. This suggests a possible phase transition from the fluid to the
Wigner crystal in the QHE system.
%---

%---
In these treatments the matrix describing particle or hole is infinite
dimensional, but Polychronakos has proposed another model of QHE
using finite matrix~\cite{QHP}.
%---

%---
As was stated above, we have to introduce two matrix model in some case, 
or equivalently $U(1)$ gauge field should be duplicated there.  In this respect it is interesting to study $U(1) \times U(1)$ 
non-commutative CS theory, or more generally non-commutative $U(2)$ CS theory. 
Singular gauge transformations in $U(2)$ model may be related to  the solitonic 
states such as quasi-particle, quasi-hole and exciton.
%---

%---
The purpose of this paper is to generalize the method developed by Polychronakos 
in studying singular gauge transformation in non-commutative gauge theory. 
His method uses the Seiberg-Witten map~\cite{SW}.  This mapping relates gauge fields 
and gauge transformations having different non-commutative parameters $\theta$. 
In QHE the non-commmutative parameter $\theta$ is inversely proportional to the 
filling fraction $\nu$ as follows:
\begin{equation}
\theta=\frac{1}{2\pi \rho}=\frac{1}{eB\nu},
\end{equation}
where $\rho$, $e$, and $B$ are, respectively, the density, the charge
of electron, and the magnetic field applied from outside.
Therefore, if the magnetic field is kept constant, Seiberg-Witten map may describe 
the change of quasi-particle, quasi-hole, and exciton states under the change of 
the filling fraction $\nu$. Therefore, the method may be useful to study the 
transition of states in Quantum Hall effect occuring when the filling fraction 
is changed.

We first review the work by Porychronakos on singular gauge
transformation in non-commutative CS theory with $U(1)$ gauge group.
Next,
we generalize it to the same theory with $U(2)$ gauge group.  Discussions are
prepared finally.
%%%%%%%%%%%%%%%%%  SECTION 2 %%%%%%%%%%%%%%%%%%%%%%%%%%%%%%%%%%%%%%%%%%
\section{Operator formalism of Seiberg-Witten map and singular gauge transformation}
%%%%%%%%%%%%%%%%%%%%%%%%%%%%%%%%%%%%%%%%%%%%%%%%%%%%%%%%%%%%%%%%%%%%%%
First, we give a brief explanation on Seiberg-Witten (SW) map and
Seiberg-Witten (SW) equation~\cite{SW}.  The SW map is the expression of
gauge field $\hat{A}$ and gauge transformation parameter $\hat{\lambda}$
in a non-commutative gauge theory in terms of those, $A$ and $\lambda$,
in a commutative gauge theory, namely,
%---
\begin{eqnarray}
\hat{A}&=&\hat{A}(A), \\
\hat{\lambda}&=&\hat{\lambda}(\lambda).
\end{eqnarray}
%---
Then, the following consistency condition is naturally imposed:
%---
\begin{equation}
\hat{A}(A)+\hat{\delta}_{\hat{\lambda}}\hat{A}(A)
 =\hat{A}(A+\delta_{\lambda}A),
 \label{SW}
\end{equation}
%---
where $\hat{\delta}_{\hat{\lambda}}$ and  $\delta_{\lambda}$ are gauge
transformations of the non-commutative theory and the commutative theory, respectively. 
From this condition, SW map is determined.
%---

%---
The operator formalism of SW map is given originally by Kraus and Shigemori~\cite{SWO}. 
Polychronakos modified it so that the SW map may preserve the hermiticity~\cite{SWOC}, 
which we will review in the following.
%---

%---
We consider D dimensional space, in which $2n$ dimensions are non-commutative 
coordinates, satisfying
%---
\begin{equation}
 [x^\alpha,x^\beta]=i\theta^{\alpha\beta},
  \label{u11}
\end{equation}
%---
while the remaining $D-2n$ dimensions  are commutative coordinates. 
The middle Greece indices,
$(\mu,\nu,\cdots=1,\cdots D)$ denote all dimensions, early
Greece indices, $(\alpha,\beta,\cdots=1,\cdots 2n)$ denote
non-commutative dimensions, and $(i,j,\cdots=2n+1,\cdots D)$ denote
commutative dimensions.
Then, consider $U(N)$ gauge theory, where the gauge field $A_\mu$ is
$N\times N$ hermitian matrices.
%---

%---
We will start to explain the operator formalism of SW equation.
If we obtain SW maps using Eq.(\ref{SW}) for two different theories
with two different non-commutative parameters differing infinitesimally by 
$\delta\theta^{\alpha\beta}$, then the non-commutative gauge fields 
$A(\theta+\delta\theta)$ and $A(\theta)$ in two differnt theories are related, 
giving
%-----
\begin{equation}
 \delta
A_\mu=A_\mu(\theta+\delta\theta)-A_\mu(\theta)
     =\frac{1}{4}\delta\theta^{\alpha\beta}
      \{
        A_\alpha,\partial_\beta A_\mu+F_{\beta\mu}
        \},
    \label{u12}
\end{equation}
%---
where the products of fields are understood to be the usual star products,
and the field strength is given as usual,
%---
\begin{equation}
 F_{\mu\nu}=\partial_{\mu}A_\nu-\partial_{\nu}A_\mu-i[A_\mu,A_\nu].
  \label{u13}
\end{equation}
%---
From (\ref{u12}), we can derive the changes of $F_{\mu\nu}$ and gauge transformation 
parameter $\lambda$ similarly as in $A_\mu$.  
These equations giving variation of fields and parameters under the infinitesimal change 
of $\theta$ is called SW equations.
%---

%---
In the following, we always work with the covariant derivatives, 
$D_\mu=i\partial_\mu+A_\mu$. The derivative in the commutative direction is as usual, 
but that in the non-commutative direction should be treated carefully.
%---

%---
First, we rewite the simple derivative in the non-commutative direction as
%---
\begin{eqnarray}
 i\partial_\alpha=\omega_{\alpha\beta}x^\beta ,
 \label{u14}
\end{eqnarray}
%---
This can be understood easily, since
%---
\begin{eqnarray}
 [i\partial_\alpha,x^\beta]=i\delta^\beta_\alpha,
 \label{u16}
\end{eqnarray}
holds if $\omega_{\alpha\beta}$  is the inverse 2-form  of
$\omega_{\alpha\beta}$, namely,
%---
\begin{eqnarray}
 \omega_{\alpha\beta}\theta^{\beta\gamma}=\delta^\gamma_\alpha.
 \label{u15}
\end{eqnarray}
%---
Important point here is that the simple derivative, $i\partial_\alpha$,
is not a usual commutative operator but is the non-commutative operator.
%---

%---
Hence, the covariant derivative in the non-commutative direction reads
%---
\begin{eqnarray}
 D_\alpha=i\partial_\alpha+A_\alpha=\omega_{\alpha\beta}x^\beta+A_\alpha
\quad,
 \label{u17}
\end{eqnarray}
%---
and the field strength satisfies the following equation  being different from the 
usual one,
%---
\begin{eqnarray}
 iF_{\alpha\beta}=[D_\alpha,D_\beta]+i\omega_{\alpha\beta}
  \label{u18}
\end{eqnarray}
%---
Now, SW equation for the covariant derivative needs
%---
\begin{eqnarray}
 \delta D_\mu=-\omega_{\mu\nu}\delta\theta^{\nu\rho}D_\rho
 +\frac{i}{4}\delta\theta^{\alpha\beta}\{D_\alpha ,[D_\beta ,D_\mu]\}
 +i[{\delta}G,D_\mu],
  \label{u113}
\end{eqnarray}
%%---
where ${\delta}G$ is an non-covariant object, given by
%---
\begin{eqnarray}
 {\delta}G(D)=\frac{1}{4}\delta\theta^{\alpha\beta}
  \{i\partial_\alpha,D_\beta\}.
  \label{u114}
\end{eqnarray}
%---
The last term in (\ref{u113}) with ${\delta}G(D)$ can be regarded as an infinitesimal 
gauge transformation.  
As is easily understood, the difference of two covariant objects (covariant derivatives) 
defined in two different theories are not gauge invariant in both theories, 
so that the difference may be generated by a gauge transformation, ${\delta}G(D)$, 
of the covariant object (covariant derivative) in one of two theories.
%---

%---
Next, we study singular gauge transformation and its topology.
Gauge transformation of the covariant derivative by a unitary
transformation $U(\theta)$ is as follows:
%---
\begin{eqnarray}
D_\alpha(\theta)=U(\theta)^{-1}D^{(0)}_{\alpha}(\theta)U(\theta),
\end{eqnarray}
%---
where we write $\theta$ explicitly, in order to specify the theory.
The variation of this equation for an infinitesinal change of $\theta$
is
%---
\begin{eqnarray}
 {\delta}D_\mu(\theta)
 =U(\theta)^{-1}({\delta}D^{(0)}_{\mu}(\theta))U(\theta)
  +({\delta}U(\theta)^{-1})D^{(0)}_{\mu}(\theta)U(\theta)
  +U(\theta)^{-1}D^{(0)}_{\mu}(\theta)({\delta}U(\theta))  
\label{eq16}
\end{eqnarray}
%---
Applying the SW equation for covariant derivatives in (\ref{eq16}),
we obtain the following result:
%---
\begin{eqnarray}
iU(\theta)^{-1}{\delta}U(\theta)
={\delta}G\left(U(\theta)^{-1}D(\theta)U(\theta)\right)
 -U(\theta)^{-1}{\delta}G(U(\theta))U(\theta).
 \label{original}
\end{eqnarray}
%---
If we start from the vacuum having $A^{(0)}=0$, we have
%---
\begin{eqnarray}
D^{(0)}_{\mu}(\theta)&=&i\partial_{\mu},  
 \label{eq.18} \\
D_{\mu}(\theta)&=&U(\theta)^{-1}i\partial_{\mu}U(\theta)=A_{\mu}(\theta)^\prime
,  
\label{eq.19}
\end{eqnarray}
%---
so that the covariant derivative $D_{\mu}(\theta)$ is identical to a solitonic
gauge field obtained from the vacuum by a singular gauge transformation $U(\theta)$ 
in non-commutative gauge theory.
%---

%---
Here, we review an example of the singular gauge transformation given by
Polychronakos in 1+2 dimensional $U(1)$ non-commutative gauge theory in which 
we have a commutative time, $t$, and non-commutative space coordinates, $x^1$
and $x^2$. To represent the non-commutativity of the space it is useful
to introduce the annihilation and creation operators, $a$ and $a^{\dagger}$, by
%---
\begin{eqnarray}
a=\frac{x^1+ix^2}{\sqrt{{\mathstrut}2\theta}},\qquad [a,a^{\dagger}]=1,
  \label{u133}
\end{eqnarray}
%---
and prepare a vacuum $|0\rangle$ satisfying $a|0\rangle=0$ and the Fock states 
$|n\rangle (n=1,2, \dots)$.
Then the covariant coordinates $X^1$ and $X^2$ can be defined by
%---
\begin{eqnarray}
 Z=\frac{X^1+iX^2}{\sqrt{{\mathstrut}2\theta}}=U^{-1}aU,
  \label{u134}
\end{eqnarray}
%---
which can be understood from Eqs. (\ref{eq.18}) and (\ref{eq.19}), 
since the covariant coordinates introduced here are the conjugate variables of 
the covariant derivarives.
%---

%---
Rewriting (\ref{u114}) in terms of covariant coordinates, we obtain
\begin{eqnarray}
 i{\delta}G=\frac{i}{4}\delta\omega_{\alpha\beta}\{x^\alpha,X^\beta \}
          =\frac{\delta\theta}{4\theta}
           (\{a^\dagger,Z\}-\{a,Z^{\dagger}\}).
    \label{u135}
\end{eqnarray}
However, there is a freedom to modify the gauge transformation ${\delta}G$ 
within the admissible gauge transformation~\cite{QHP}, so we can use instead of 
${\delta}G$ a simpler gauge transformation ${\delta}G^{\prime}$ given by
%---
\begin{eqnarray}
i{\delta}G^{\prime}=\frac{\delta\theta}{2\theta}(a^{\dagger}Z-Z^{\dagger}a).
 \label{deltaG'}
\end{eqnarray}
%---
If we assume that the unitary transformation $U(\theta)$ takes the following form,
%---
\begin{eqnarray}
 U(t; \theta) =\sum^\infty_{n=0}e^{i\phi_n(t; \theta)}|n{\rangle}{\langle}n|,
  \label{u137}
\end{eqnarray}
%---
then we have the following  differential eqaution from (\ref{original}):
\begin{eqnarray}
 \delta\phi_n=\frac{\delta\theta}{\theta}n\sin(\phi_n-\phi_{n-1})\qquad
.
  \label{u138}
\end{eqnarray}
We can solve this equation, starting from a non-commutative theory with 
non-commutaitve parameter $\theta=\theta_0$, under the following boundary conditons
%---
\begin{eqnarray}
& &\phi_0(t=+\infty; \theta_{0})-\phi_0(t=-\infty; \theta_{0})=2\pi, \label{u130-1} \\
& &\phi_n(t; \theta_{0})=0~~ \mbox{for}~~(n=1, 2, \dots).
 \label{u130-2}
\end{eqnarray}
 %---
The dependence of $\phi$ on $n$ gives the dependence of $\phi$ on the spacial radius 
$r=\sqrt{(x^1)^2+(x^2)^2}$, since we have approximately 
$r\approx \sqrt{\mathstrut 2n\theta}$. 
Eq.(\ref{u130-1}) means that a topological excitation is created around $r=0$.
%---

%---
The solution of the above differential equation shows that the gauge transformation viewed from the commutative theory with $\theta=0$ gives  a singular behavior when $\phi$ approaches the value $\pi$ at the origin $r=0$ ~\cite{SWOC}.  Namely, the value of $\phi$ at $r=0$ increases in time and attains $\pi$, but then suddenly jumps to $-\pi$ and increases again in time.  There appears a kink at $r=0$ as a function of $t$, and spacial profile near the kink is a pulse with height $\pi$ and a pulse with height $-\pi$ before and after the time when $\phi$ crosses the kink position. This singular behavior is a characteristic of the solitonic solution in non-commutative gauge theory. Figure, (1), (2), (3), and (4) show the spacial profiles of the phase $\phi$ of singular gauge transformation at time $t_1, t_2, t_3,$ and $t_4\;(t_1<t_2<t_3<t_4)$, respectively.
%%%---ここに図を入れる---%%%
\begin{figure}[ht]
\begin{center}
\includegraphics[width=4cm,clip]{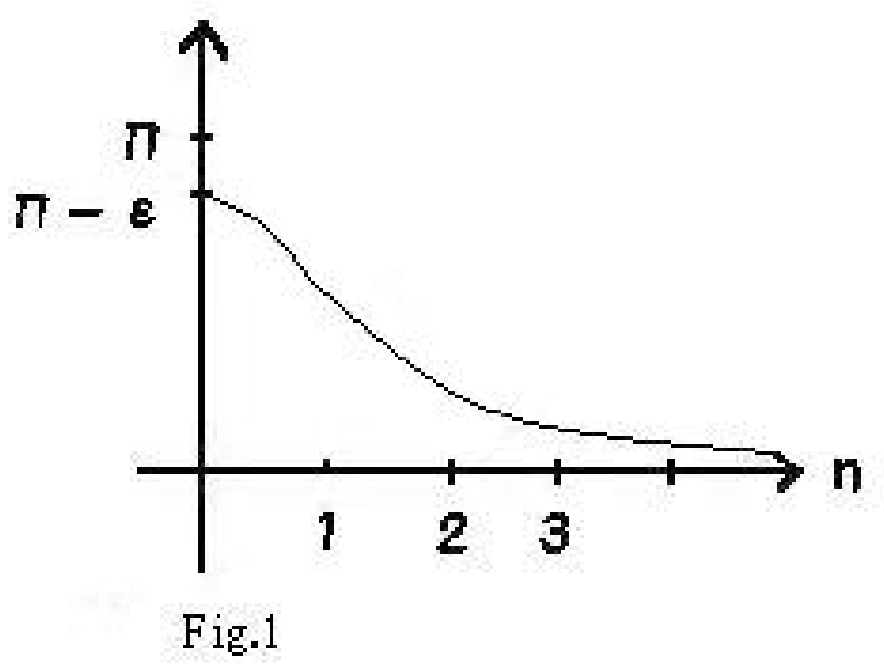}
\includegraphics[width=4cm,clip]{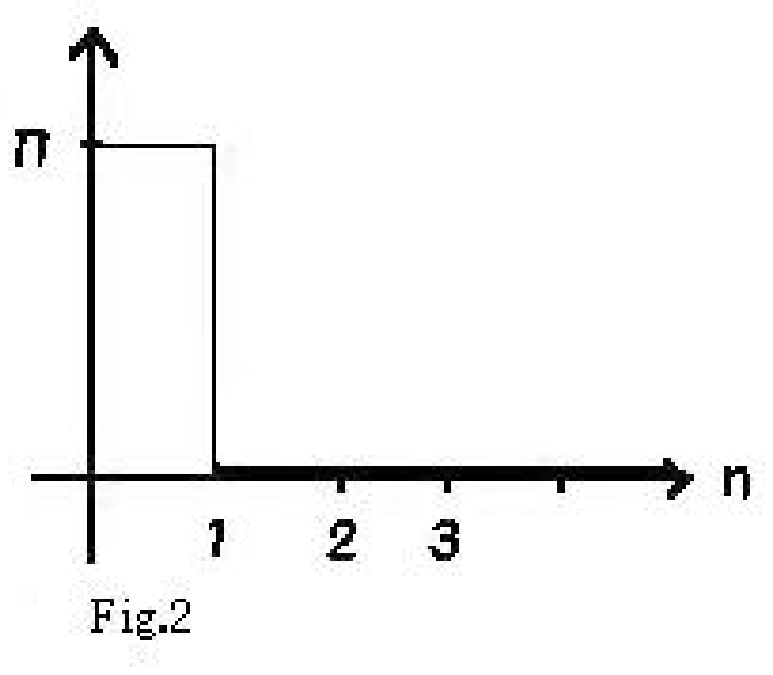}
\includegraphics[width=4cm,clip]{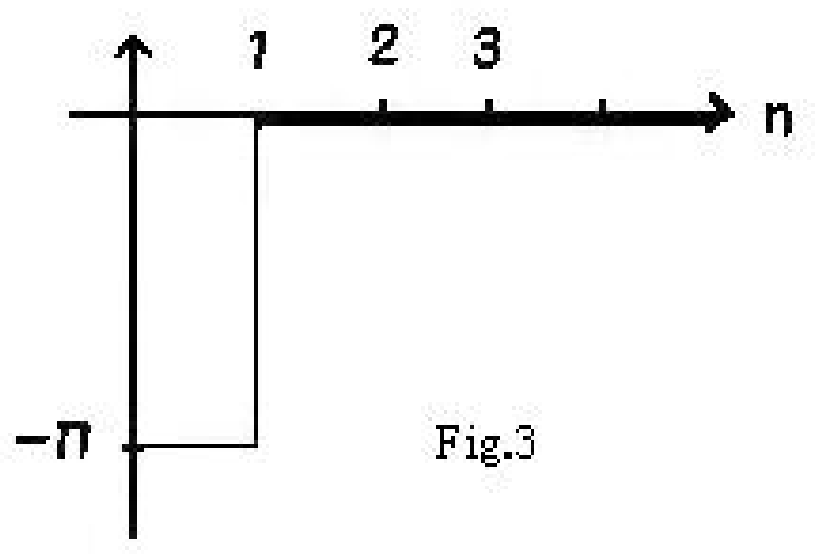}
\includegraphics[width=4cm,clip]{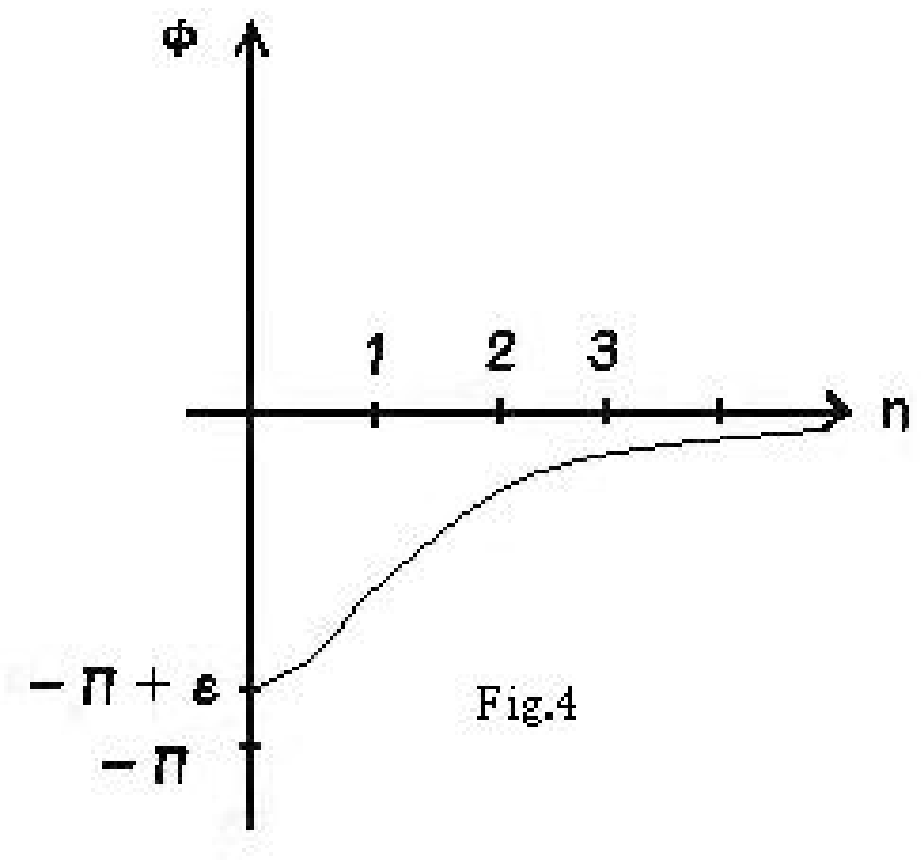}
\caption{Figs. (1), (2), (3), and (4) depict the spacial profile ($n$ dependence) of the phase $\phi$ at $t=t_1, \;t_2, \; t_3,$ and $t_4$ $(t_1 \; <\; t_2 \; < \; t_3 \;<\; t_4)$, respectively.
} 
\end{center}
\end{figure}
%%%%%%%%%%%%%%%%%%%%%%
%%%%%%%%%%%%%%%%%%%%%%
%%%%%%%%%%%%%   SECTION 3 %%%%%%%%%%%%%%%%%%%%%%%%%%%%%
\section{Generalizaion to U(2) gauge group}
%%%%%%%%%%%%%%%%%%%%%%%%%%%%%%%%%%%%%%%%%%%%%%%%%%%%%%
In this section, we will generalize the prescription of studying singular gauge transformation to the non-commutative gauge theory with $U(2)$ group.
The gauge tranformation should satisfy the SW equation (\ref{original}) under the cahnge of the non-commutative parameter$\theta$,
%---
\begin{eqnarray*}
 iU^{-1}{\delta}U={\delta}G^{\prime}(U^{-1}DU)-U^{-1}{\delta}G^{\prime}(D)U,
\end{eqnarray*}
%---
where we have adopted the simpler choice of ${\delta}G^\prime$ in (\ref{deltaG'}).
Then, we obtain
%---
\begin{eqnarray}
 iU^{-1}{\delta}U
 &=&\frac{\delta\theta}{2i\theta}(a^{\dagger}Z-Z^{\dagger}a)
  \nonumber\\
&=&\frac{\delta\theta}{2i\theta}(a^{\dagger}U^{-1}aU-U^{-1}a^{\dagger}Ua),
\end{eqnarray}
%---
where we have used ${\delta}G^{\prime}(D)=0$.
%---

%---
Now, the SW equation is reduced to
%---
\begin{equation}
{\delta}U_{n}^{\dagger}U_{n}=\frac{\delta\theta}{2\theta} n (U_{n-1}^{\dagger}U_{n}-U_{n}^{\dagger}U_{n-1}).
\end{equation}
%---
The U(2) gauge transformation is given by
%---U(2)---
\begin{eqnarray}
 U(\theta)&=& \sum^{\infty}_{n=0} U_{n} |n{\rangle}{\langle}n|,
 \\
U_n(\theta)
 &=&  \sum^{\infty}_{n=0} e^{i\phi_n}
  \left(
   \begin{array}{cc}
    {\alpha}_n & {\beta}_n\\
    -{\beta}_n^{\ast} & {\alpha}_n^{\ast}
    \end{array}
   \right),
 \label{m2}
\end{eqnarray}
%---
where $|\alpha_n|^2+|\beta_n|^2=1$, and therefore
$\delta\alpha_n\alpha_n^{\ast}+\alpha_n\delta\alpha_n^{\ast}+\delta\beta_n\beta_n^{\ast}+\beta_n\delta\beta_n^{\ast}=0$ should hold for the $U(2)$ transformation.
%---

%---
%%%%%---以下の式の第2式第3式の符号をチェックする必要あり----%%%%
Then, the SW equations read
\begin{eqnarray}
 & &\delta\phi_n= -\frac{\delta\theta}{2\theta}n
    \sin(\phi_n-\phi_{n-1})
    \left(\alpha_{n}^{\ast}\alpha_{n-1}+\beta_{n}\beta_{n-1}^{\ast}+ (h.c.) \right), \\
& &\delta\alpha_{n}=-\frac{\delta\theta}{2\theta}n 2\cos(\phi_{n}-\phi_{n-1})(\alpha_{n}-\alpha_{n-1}),\\
& &\delta\beta_{n}=-\frac{\delta\theta}{2\theta}n 2\cos(\phi_{n}-\phi_{n-1})(\beta_{n}-\beta_{n-1}).
 \label{m3}
\end{eqnarray}
%---
%%%%%%----
If we express $\alpha_n$ and $\beta_n$ in terms of three phases, $\vartheta_n$, $\psi_n$, and $\chi_n$, the $U(2)$ property is manifestly guranteed.  That is, we use the following parametrization:
%---
\begin{eqnarray}
 \alpha_n \equiv e^{i\psi_n}\sin \vartheta_n, \\
 \beta_n \equiv e^{i\chi_n}\cos \vartheta_n.
\end{eqnarray}
%---
Now, we have the SW equations as follows:
%---
\begin{eqnarray}
%%%---\phi---%%%
\delta\phi_n&=&-\frac{\delta\theta}{\theta}n\sin(\phi_n-\phi_{n-1})
 \nonumber\\
 &\times&
 \left(
  \cos\vartheta_{n}\cos\vartheta_{n-1}\cos(\chi_n-\chi_{n-1})
  +\sin\vartheta_{n}\sin\vartheta_{n-1}\cos(\psi_n-\psi_{n-1})
 \right),
\label{phi}\\
%%--\theta--%%%
\delta\vartheta_n &=&-\frac{\delta\theta}{\theta}n\cos(\phi_n-\phi_{n-1})
 \nonumber\\
 &\times&
  \left(
   \sin\vartheta_{n}\cos\vartheta_{n-1}\cos(\chi_n-\chi_{n-1})
   -\cos\vartheta_{n}\sin\vartheta_{n-1}\cos(\psi_n-\psi_{n-1})\right),
\label{theta}\\
%---\psi--%%%
\delta\psi_n&=& -\frac{\delta\theta}{\theta}n\cos(\phi_n-\phi_{n-1})
 \left(
  \frac{\sin\vartheta_{n-1}}{\sin\vartheta_{n}}
 \right)
  \sin(\psi_n-\psi_{n-1}), 
\label{psi}\\
%%---\chi---%%%
\delta\chi_n&=& -\frac{\delta\theta}{\theta}n\cos(\phi_n-\phi_{n-1})
 \left(
  \frac{\cos\vartheta_{n-1}}{\cos\vartheta_{n}}
 \right)
 \sin(\chi_n-\chi_{n-1}).
\label{xi}
\end{eqnarray}
%%---general discussion, before taking the limit of n to \infty---%%%%
In general, $\phi_0, \vartheta_0, \psi_0,$ and  $\xi_0$ for $n=0$ do not change during 
the change of $\theta$ moving from the original $\theta_0$ to zero. 
Therefore these phases keep the profiles in the original theory.
The phases for $n>0$ may change, when $\theta$ is reduced towards zero,
namely $\frac{d\theta}{\theta}<0$.
In the region well apart from the location of pulses, 
we can consider that the difference of phases at $n$ and $n-1$ are not large, 
so that we can set the following approximation:
%---
\begin{eqnarray}
 \left\{
  \begin{array}{c}
  \phi_n \simeq \phi_{n-1}\\
  \vartheta_n \simeq \vartheta_{n-1}\\
  \psi_n \simeq \psi_{n-1}\\
  \chi_n \simeq \chi_{n-1}\\
\end{array}
 \right.
\end{eqnarray}
%---
Therefore in this region, we have approximately
 \begin{eqnarray}
\delta\phi_n &\simeq& -\frac{\delta\theta}{\theta}n \sin(\phi_n-\phi_{n-1}),
\label{phi'}\\
\delta\vartheta_n &\simeq& -\frac{\delta\theta}{\theta}n \sin(\vartheta_n-\vartheta_{n-1}), 
\label{theta'} \\
\delta\psi_n &\simeq& -\frac{\delta\theta}{\theta}n\sin(\psi_n-\psi_{n-1}),
 \label{psi'} \\
\delta\chi_n&\simeq& -\frac{\delta\theta}{\theta}n\sin(\chi_n-\chi_{n-1}).
\label{xi'}
\end{eqnarray}
%---
These four phases satisfy approximately the same SW equations in general. Hence we can follow the discussion which is given originally by Polychronakos and is reviewed in the last section. If the phases approach $\pi$ from below, the difference of phases between $n$ and $n-1$ increase when $\theta$ decreases, while the phases exceed  $\pi$, the difference of the phases decrease (because of the sign change of the sine function).  However, all the phases should be zero asymptotically for $n \to \infty$, or $r \to \infty$. Therefore, the region of phases exceeding $\pi$ should be understood as $-\pi$ by periodicity, so that even if the difference of phases between $n$ and $n-1$ decrease, the phases can be asymptotically zero for $n\to\infty$.
Roughly speaking, the asymptotic behavior of phases for $n \to \infty$ can be expressed in terms of the differential equations:
%---
\begin{eqnarray}
\left(
\frac{\partial}{\partial \ln\theta} \pm \frac{\partial}{\partial \ln n}
\right) 
\left(\phi, \vartheta, \psi, \xi \right)=0,
\end{eqnarray}
%---
where $\pm$ sign is the remnant of the original sign of the sine function. The + sign describes the region of phase difference from $0$ to $\pi$, and the - sign describes the region of phase difference from $-\pi$ to $0$.
From this differential equation, all the phases are asymptotically a function of  the ratio $\frac{n}{\theta}$ if the phase difference is from $0$ to $\pi$, while they are function of $n\theta$ if the phase difference is from $-\pi$ to $0$.  This gives the quantitative behavior of the change of phase when $\theta$ decreases.
%---

%---
In the $U(2)$ case, we have four phases, one of which $\phi$ is the $U(1)$ phase, but the remaining three are $SU(2)$ phases.  The same discussion can be applied for three $SU(2)$ phases as in the $U(1)$ phase. Therefore if we denote these phases generally $\Phi$. Then, if $\Phi$ is  topologically non-trivial,
\begin{eqnarray}
& &
\Phi_{0} (t=+\infty; \theta_{0})-\Phi_{0} (t=-\infty; \theta_{0})= 2\pi, \\
& &
\Phi_{n} (t; \theta_{0}) = 0 \qquad (n=1,2,\cdots),
\end{eqnarray}
we have the same singular behavior,  having kink and pulse for the $SU(2)$ phases.  Here one commutative dimension $t$ and two non-commutative dimensions $x^1$ and $x^2$ are treated separately, so that only $U(1)$-like singular behavior is obtained.  Relevant topology here is $\pi_1(U(2))$. To obtain singular behavior specific to $U(2)$, being relevant to the topology of $\pi_2(U(2)/Z_2)$, we have to consider a more general dependence of the phases on $t, x^1$ and $x^2$ which is not discussed in this paper.
%%%----
%%%%%%%% SECTION 4 %%%%%%%%%%%%%%%%%%%%%%%%%%%%%%%%%%%%%%%%
\section{Conclusion}
%%%%%%%%%%%%%%%%%%%%%%%%%%%%%%%%%%%%%%%%%%%%%%%%%%%%%%%%%%%

%%%%-----------------
We have studied singular gauge transformations in the  non-commutative U(2) gauge theories, by generalizing the method proposed by Polychronakos~\cite{SWOC} to study the singular gauge transformation in non-commutative U(1) gauge theories. The space-time dimension of the theory is three in which time coordinate is commutative, but two spacial coordinates are non-commutative with each other.  A typical example is the 1+2 dimensional non-commutative Chern-Simons gauge theory applicable to Quantum Hall effects.
%----

%----
The method uses the operator formalism of Seiberg-Witten map~\cite{SW} which connects different theories with different non-comutative parameters $\theta$.  Using this map, the gauge transformation in the non-comutative theory can be viewed from the corresponding commutative theory equivalent to the non commutative theory. In $U(2)$ gauge transformation we have four phases. SW-equations for these four phases are derived explicitly. From these equations, we obtain a similar singular behavior for these phases as is depicted in Figs. (1), (2), (3), and (4).
%---

%---
In order to study the exciton in Quantum Hall effects(QHE), it may be necessary to introduce two matrix model.  One matrix model of Chern-Simons(CS) type  and the non-commutative  Chern-Simons(CS) gauge theory with $U(1)$ group describe equivalently QHE, so that the study of singular gauge transformation in non-commutative $U(2)$ CS gauge theory may be useful for the study of QHE using two matrix model.
%---

%---
It is also interesting to connect the change of the non-commutative parameter $\theta$ by Seiberg-Witten map to the change of filling fraction $\nu$, and to study the phase structure of QHE in terms of the filling fraction $\nu$.
%%%----------
%--------------------------------
%
%         acknowledgments
%
%--------------------------------
\section*{Acknowledgment}
The authors are grateful to Taro Tani for valuable discussions, and to Midori Obara for technical supports.
%%%%%%%%%%%%%%%%%%%%%%%  reference      %%%%%%%%%%%%%%%%%%%%%%%%%

%%%%%%%%%%%
\end{document}